# Development of high-performance iron-based superconducting wires and tapes

Yanwei Ma*

Key laboratory of applied superconductivity, Institute of Electrical Engineering, Chinese Academy of Sciences, Beijing 100190, China

**Abstract:**

Conventional powder-in-tube (PIT) method has been the most effective technique for fabricating iron-based superconducting wires and tapes. Tremendous advances have been made during the last few years, especially for 122 family pnictide tapes. Here we review some of the most recent and significant developments in making high-performance iron-based tapes by the *ex-situ* PIT process, paying particular attention to several fabrication techniques to realize high-field $J_c$ performance in terms of increase of core density and improvement of texture. At 4.2 K, the practical level transport $J_c$ up to 0.12 MA/cm$^2$ in 10 T and 0.1 MA/cm$^2$ in 14 T have been achieved in the K-doped 122/Ag tapes. As for multifilamentary 122 iron-based wires and tapes, the highest $J_c$ values reached so far are 61 kA/cm$^2$ and 35 kA/cm$^2$ at 4.2 K and 10 T, respectively for 7- and 19-core Sr-122 tapes. Recently, high $J_c$ Cu-cladded and stainless steel/Ag double-sheathed 122 tapes have also been produced in order to improve either mechanical properties or thermal stability. More importantly, the scalable rolling process has been used for the first time to demonstrate high $J_c$ values in 122 conductor tapes of 11 m in length.

---

* Author to whom correspondence should be addressed; E-mail: ywma@mail.iee.ac.cn

**Contents**

# 1. Introduction

The discovery of superconductivity in iron-pnictides in 2008 [1] has promoted a new interest in the area of fundamental and applied research on superconducting materials. Soon after, many other series of new compounds have been identified. Among them, SmOFeAs ($T_c$= 55 K, 1111 type) and Sr/BaKFeAs ($T_c$= 38 K, 122 type) seemed the most relevant for applications due to higher transition temperature. With the highest $T_c$, pnictides have other advantages of ultrahigh critical fields ($H_{c2}$>50 T at 20 K), large critical current densities $J_c$ of the order of $10^6$ A/cm$^2$ in thin films and low anisotropy ($\gamma$<2 for 122) [2-4]. These are properties that render pnictides very promising for potential applications. Moreover, it is well understood that the weak link effect in iron pnictides is not as heavy as in YBCO, since the critical angle that begins to affect supercurrents is up to 9º for pnictides, which is substantially larger than 3~5º of YBCO [5]. This means that a scalable fabrication process can be applied for the fabrication of pnictide wires and tapes [6]. On the other hand, as a final consideration for the practical use of superconducting wires, the current density calculated over the full conductor cross-section, or engineering current density Je, needs to be at least $10^4$ A/cm$^2$ under operating conditions.

The powder-in-tube (PIT) method is most widely used for the fabrication of 1111, 11 (FeSe) and 122-type superconducting wires and tapes using various sheath materials such as Fe, Ag, Nb and Ta [6-8]. However, the reported transport $J_c$ at an early stage was very low due to the presence of a lot of impurities, cracks and voids within the superconducting core as well as the misorientation between the grains. The progress in the area of iron-based wires and tapes has been the result of successfully overcoming a series of major difficulties including the reaction between the superconductor core and the tube [9], the low quality of the starting powder [10], and poor grain connectivity [11, 12]. In recent years, enormous efforts directed towards fabrication of useable iron-based wires and tapes have been reported, including chemical doping [13], texturing [14, 15], the hot isostatic pressing [16], cycles of rolling plus intermediate annealing [17] and cold pressing deformation techniques [18,19], to improve the transport critical current density $J_c$ in high magnetic fields. Recently, our group reported an alternative approach, that is, hot pressing as a final sintering, for the fabrication of 122 superconducting tapes with large transport critical current density. At 4.2 K, the $J_c$ values of these pressed tapes exceeded $10^5$ A/cm$^2$ at fields as high as 10 T [20], which has surpassed the threshold for practical application.

In this paper, we review key properties and technical progress in developing high performance iron-based superconducting materials during the last few years. The article is aimed at providing a snapshot of these recent developments and a sense of their significance for high-field applications.

## 2. The *ex-situ* PIT process

The PIT process has advantage of the low material costs and the deformation techniques used are very simple. There are two PIT methods. One is an in-situ method, in which a powder mixture of precursor of 122 phase is used as a starting material. The other is an ex-situ method, in which powder of 122 phase is used. Currently, the most successful approach for fabricating pnictide wires was the so-called *ex-situ* PIT method, first used for 122 type by our group [11]. This method is attractive because the fabrication process leads to fewer impurity phases as well as a high density of the superconducting core for the final wires. Figure 1 shows the schematic illustration of PIT method. In the case of Sr-122 type, Sr fillings, K pieces, and Fe and As powders are used as staring materials with the nominal composition of $Sr_{0.6}K_{0.4}Fe_2As_2$. In order to compensate for the loss of elements during the sintering, the starting materials contains 25 wt% excess K and 5 wt% excess As [21]. The starting materials are mixed and ground by ball-milling method under argon atmosphere. Then milled powders are heat treated at 900 ℃ for 35 h. The sintered bulk is ground into powders and mixed with 5 wt% Sn. It is well established now that the formation of the Sr-122 phase is enhanced by adding with Sn [22]. The Sr-122 reacted precursor is packed into silver (Ag) tubes or iron (Fe) tubes with OD 8 mm and ID 5 mm under Ar atmosphere and sealed to form billets. The billets are then swaged and drawn, flat rolled to tapes (about 0.3-0.6 mm thickness) and finally given a recovery heat treatment at 850-900 ℃ for under an Ar gas atmosphere.

## 3. Optimization of fabrication process

### 3.1 Optimal sintering process

During the deformation process, as shown in figure 1, the 122 grains are densified by the stresses introduced by cold working, cracks and defects are formed in the filaments, therefore the final heat treatment should be needed for reconnecting the grains, which usually leads to a considerable increase of the core density and the grain connectivity is strongly improved, thus causing a strong increase of $J_c$.

In order to find optimal heat conditions, Lin et al. recently studied the effects of sintering temperatures on the phase formation, resistivity transition, critical current density and microstructure of Sn-added Sr-122 tapes [22]. The sintering time was fixed at 0.5 h, and the sintering temperature was varied from 800 to 950 ℃. From the X-ray diffraction (XRD) analysis, almost all XRD peaks are attributed to Sr-122 superconducting phases, however, small amount of FeSn impurity was almost always detected in XRD patterns, and its content increased at higher sintering temperature, especially at 950 ℃.

Transition temperature $T_c$ of Sr-122 tapes increases with increasing the heat treatment temperature. For the tapes prepared at 900 ℃, the onset $T_c$ = 33 K and transition width ΔT = 1.58 K was obtained. When at 950 ℃, although the onset Tc was 33 K, broad width ΔT = 2.82 K was observed. This can be explained by more impurity phases induced by the higher temperature, as supported by the XRD data. Figure 2 shows the field dependence of transport $J_c$ at 4.2 K for the Sr-122 tapes with different sintering temperatures. Clearly, in-field $J_c$ values for samples heated at temperatures between 850 and 900 ℃ are much larger than those of tapes sintered at 800 and 950 ℃, suggesting the significant effect of sintering temperatures on the $J_c$ of Sr122 tapes. Microstructural analysis demonstrated that the sample sintered at 800 ℃ had non-uniform particle sizes and irregular morphology because this reaction temperature is 100 ℃ lower than the synthesis temperature of precursors, which is unable to completely recover the cracks and pores induced in cold deformation. As for tapes heated at 950 ℃, many black pores and white dendrite structures inside the matrix were existent, caused by the excessive loss of the volatile component of superconducting phase, such as potassium. The pores and also impurity phases resulted in a drastic $J_c$ degradation, as further supported by XRD and R-T analysis.

In summary, the sintering temperature between 850 and 900 °C is most suitable to realize high $J_c$ for 122 type wires.

**3.2 Optimal rolling process**

The typical PIT thermomechanical processing for 122 pnictide wires consists of one rolling step and one heat treatment step. The purpose of the rolling step is to densify 122 filaments with high density core and connectivity between grains is recovered after a final heat treatment. Therefore, optimization of the rolling process is also important for achieving high critical current density values through the fabrication well connected 122 crystal structure.

Thus far, Ag and Fe are two sheath materials currently used for 122 wire fabrication. Ag looks the most appropriate sheath material for 122 phases, because Ag does not react with 122 superconductor, as first demonstrated by Wang et al. [23], and soon has been confirmed by several leading groups, such as the National Institute for Materials Science, NIMS (Japan), the Florida High Field National Laboratory (US), the University of Tokyo (Japan). From the viewpoint of low cost, Fe has been found to be another candidate sheath material for 122 tapes and wires [6], however, one of the main problems encountered when producing Fe-sheathed wires or tapes is the reaction with 122 phases during the final heat treatment. Since Fe has higher hardness than Ag, deformation properties of flat rolling were quite different as shown below.

Figure 3 presents the field dependent transport $J_c$ at 4.2 K of the rolled Sr-122/Ag tapes with different thickness and ϕ1.9 mm round wires. The $J_c$ of the ϕ1.9 mm round wires in self-field is $10^4$ A/cm$^2$. However, it rapidly decreases by almost one order of magnitude when the application of external magnetic field and then shows a very weak field-dependence in the high field region. The $J_c$ of the 1.9 mm round wires at 10 T seems low, 643 A/cm$^2$, which is comparable to those reported so far for Ag-sheathed PIT 122 round wires fabricated by the conventional PIT process [11, 24]. It is noted that the transport $J_c$ drastically increases after the flat rolling, and further increases when the rolling thickness is reduced. The $J_c$ achieved a maximum value of $2.3\times10^4$ A/cm$^2$ at 10 T at the thickness of 0.3 mm of tapes. This rapid increase of $J_c$ should be attributed to the increase of the core density. The $J_c$ reproducibility is good for the tape thickness around 0.3 mm. When further rolling the tape to small thickness, the $J_c$ degradation is observed. Similar behavior of the rolled Ba-122/Ag tapes was also reported by Togano et al. [17].

On the other hand, as for the Fe-sheathed Sr-122 wires, like the case of Ag, the $J_c$ also rapidly increased by the flat-rolling process, but reaching a maximum value at the thickness of 0.6 mm as shown in figure 4, which is quite thicker than Ag sheathed tapes. Furthermore, Fe-sheathed Sr-122 tapes usually showed lower $J_c$ than in Ag-cladded ones. This is due to that the Fe sheath is quite prone to chemically react with the Sr-122 core and form the FeAs contained reaction layers at the temperature as high as 900 ℃. These thin reaction layers would consume some of the As within the filament, not only leading to an increase in porosity and impurities, thus a lowering of the transport $J_c$ value, but also making the Fe sheath very brittle, reducing the mechanical properties of conductors. Nevertheless, Fe is still a valuable alternative to Ag sheath, due to low cost material. Indeed, the finally annealing temperature may become lower than 900 ℃ after further optimization, thus this reaction layer is expected to be small. Therefore, Fe remains an interesting alternative for application.

In a word, the transport $J_c$ increases for smallest tapes thicknesses, a maximum being found for a tape thickness of about 0.3 mm and 0.6 mm for Ag and Fe sheathed 122 tapes, respectively, below which the formation of cracks in the filament limits the critical current. This can be understood that the optimum rolling thickness from different sheath materials was determined by the balance between improvement in density and initiation of microcracks which cannot be healed by final heat treatment.

## 4. Advances in improving $J_c$ of 122 wires and tapes

### 4.1 A texturing process

At an early stage of development, the transport $J_c$ of PIT processed wires was quite low due to the poor grain linkage problem [25, 26]. This might be due to the presence of numerous cracks and voids along grain boundaries and/or misorientation between grains. Soon later, it was understood that iron-pnictides also suffer from weak-link behavior, but is not as severe as in YBCO [5], e.g., a critical grain boundary (GB) angle 9º (at which the intergrain supercurrents start to degrade) is substantially larger than the 3~5º for YBCO. However, the intrinsic weak-link effect at misoriented GBs can deteriorate supercurrents regardless of the mass density of materials [27]. Thus, we may borrow some of the same texturing strategies that have been so vital for Bi-2223tapes to fabricate high-$J_c$ 122 pnictide tapes.

Important progress was achieved by Wang et al.[14], who first demonstrated that strong texture of *c*-axis alignment can be achieved by applying the flat rolling with large reduction ratio for Fe-sheathed Sr-122 wires and this is effective to improve the grain connectivity as well as the densification. The sheath material used was Fe because the heat treatment was carried out at a high temperature for a short time. It is clear that the rolling process promotes grain alignment with the *c*-axis perpendicular to the tape surface because 122 superconducting grains can be easily cleaved along the *a-b* plane, as shown in figure 5. Correspondingly, the grain boundaries with an alignment angle below 9º can become a favorable current path, accompanied by a higher $J_c$. The grain alignment also seems to play a crucial role in the connectivity. It is noted that this rolling process is quite simple, similar to that developed for large scale production of Bi-2223 superconducting tapes in the early 1990s [28].

Inspired by the above texturing process, Gao et al [15] approached the texturing strategy by lowing the sintering temperature in combination with the Sn addition and obtained remarkable high $J_c$ values (at 4.2 K and 10 T) of 17 KA/cm$^2$ in textured Sr-122 tapes. The well developed grain texture in Sr-122/Fe tapes is a main reason for the superior $J_c$ performance, since the misalignment in crystalline orientation at grain boundaries is a crucial weak-link in connectivity. The author's group [29] reported the transport $J_c$ at temperatures above 4.2 K using textured 122 wires and found that the $J_c$-H property keeps a weak magnetic field dependence at temperatures up to 20 K (see figure 6). At temperatures of 20 K, easily obtained using a cryocooler, $J_c$ reached ~10$^4$ A/cm$^2$ in self-field. Magneto-optical imaging observations further revealed significant and well distributed global $J_c$ at 20 K in our textured tapes. Those results indicate that the PIT processed 122 superconducting wires are promising for magnetic field applications at medium temperatures of cryogenic cooling or liquid hydrogen as well as in liquid helium.

Recently, author's group has fabricated Fe-sheathed Sr-122 tapes showing a substantial enhancement of $J_c$ after further optimizing the texturing process, e.g., the sausaging of the filament has been reduced by optimizing the thickness reduction between consecutive rolling steps. As shown in figure 7, the batch II tapes were found to exhibit the highest $J_c$ value of $2.4 \times 10^4 \text{A/cm}^2$ at 14 T, 4.2 K. Such high electrical performance is believed to come from an improvement of grain connectivity by reducing weak links.

More encouraging result in enhancing $J_c$ was achieved by Dong et al. [30] for Ag-clad $Ba_{0.6}K_{0.4}Fe_2As_2$ thin tapes using high quality precursor plus scalable rolling process. They fabricated Ba-122 tapes via only one cycle of drawing and flat rolling process. At 4.2 K and 10 T, transport $J_c$ of 54 kA/cm$^2$, and engineering critical current density $J_e$ of 15 kA/cm$^2$ have been achieved by this scalable rolling process (See figure 7). High purity phases and higher texture (F=0.69) are observed in these Ag-sheathed Ba-122 tapes. Here the degree of grain texture, c axis orientation factor F, was evaluated by XRD analysis using the Lotgering method [31]. This shows that besides a sufficiently high texture, the quality of the precursor used is also essential for obtaining high in-field $J_c$ samples.

Although the above 122 tapes have textured microstructure along the rolled metals, the anisotropy of $J_c$ with the applied magnetic field angle is quite small, less than 1.5, not only for Fe-clad Sr-122 tapes [15], but also for Ag-sheathed samples, e.g., the anisotropy ratio of Sr-122/Ag $\Gamma = J_c^\perp / J_c^{//} = \sim 1.28$, measured recently by Kovac group in Slovak, while the $\Gamma$ was ~1.32 in magnetic fields up to 15 T, performed by Dr. Awaji at the high field lab in Sendai. These low anisotropy ratios of our Sr-122 tapes are surely beneficial to practical applications. In addition to a high $J_c$, the process for fabricating wires should be simple and easy to scale up to long length km wire production, requiring a minimum of sophisticated equipment or processes. Clearly, the above-mentioned texturing or rolling is a scalable and practical process for making long and continuous lengths of pnictide conductor.

**4.2 Densification effects**

Densification is another dominant factor that determines the critical current performance of the present PIT pnictide wire. Higher $J_c$ in wire requires a high mass density of the superconducting filaments simply owing to the enhancement of the current flow path [32]. If additional methods are applied to PIT wires to increase the mass density, the values of $J_c$ can be enhanced. Weiss *et al*. achieved nearly 100% dense core inside Cu/Ag/Ba-122 round wires by applying a hot isostatic press (HIP) synthesis under 192 MPa of pressure at 600°C for 20 hours and thus the resulted large $J_c$ (at 4.2 K) of 100 kA/cm$^2$ in self field and 9 kA/cm$^2$ at 10 T [16]. The same level $J_c$ was recently obtained in a Sr-122/Ag round wires by Pyon et al. [33]. They hot isostatic-pressed Sr-122 wires at 700 °C for 4 h in Ar atmosphere under pressure of 120 MPa. The results indicate the importance of the densification of the 122-type core.

In the course of developing PIT Bi-2223 tapes, it was reported that cold pressing is more effective for the enhancement of transport $J_c$ than flat rolling [34, 35]. The advantages of cold pressed tapes showed a higher core density and a crack direction formed longitudinally along the tape, in contrast to the rolled tape. Recently, cold uniaxial pressing was first introduced by Yao et al. [18] as a route for enhancing the filament mass density for 122 type wires. The as-drawn Fe-sheathed Sr-122 wires were cut into short samples, and then directly pressed into tapes with maximal pressures of 0.6, 1.0, 1.4 and 1.8 GPa, respectively. It is found that the pressing can improve the mass density and induce c-axis texture in the Sr-122 core. The transport current density is linearly improved (maximum $J_c$ = 16.8 kA/cm$^2$ at 4.2 K in self field) until the applied pressure comes up to 1.4 GPa, and then becomes saturated at 1.8 GPa.

Just like the case of PIT Bi2223, it is interesting to follow the development of $J_c$, with the number of annealing, rolling and pressing cycles. Togano et al. [36] reported that the $J_c$ of Ba-122/Ag tapes can be much enhanced by applying a combined process of flat rolling and uniaxial pressing (maximum $J_c$ = 21 kA/cm$^2$ at 4.2 K in 10 T) to achieve denser structure in the 122 filament. The tapes were prepared by cycles of rolling and intermediate annealing, and pressing followed by the final heat treatment for sintering. Thus, the large $J_c$ enhancement of 122 tape by the cold pressing can be explained by the improved connectivity and fewer cracks running transverse to the tape length.

We fabricated Ag-sheathed Sr-122 tapes and tried to improve the $J_c$ properties by controlling the rolling, sintering and pressing steps through the improvement of the filament density and the crystal alignment of the 122 phase. As shown in figure 8, among these processes, they found that the pressing pressure exerted on the tapes between the heat treatments has a strong influence on the transport properties: a maximum $J_c$ (4.2 K, 10 T) value of 48 kA/cm$^2$ on pressed Sr-122 tapes was found for a pressing pressure of 1.0 GPa. At higher pressures, the formation of cracks in the filament limits the critical current. The result shows that the pressing is important for achieving high $J_c$ values through the formation of densely and well connected 122 crystal structure. A better enhancement of a repeated rolling and cold pressing process under high pressure ∼2GPa was also observed by Gao et al. [19] for Ba-122/Ag tapes. The transport $J_c$ reached 86 kA/cm$^2$ at 10 T and 4.2 K. This implies that high densification by rolling the wire several times and/or changing the pressure during pressing deformation process is effective in enhancing $J_c$. However, cold pressing always causes fatal micro-cracks inside the superconducting core, which cannot be healed by a subsequent heat treatment.

More recently, a significant breakthrough in enhancing $J_c$ of 122 type wires was achieved using hot pressing instead of usual heat treatment by our group [20, 37, 38]. The data on Sr-122 wires show a substantial increase of $J_c$ after hot pressing. As shown in figure 9, an improvement of $J_c$ by factor 3.4 at 4.2 K was obtained upon pressing. The transport $J_c$ of hot pressed Sr-122 tapes was over $10^5$ A/cm$^2$ in 10 T at 4.2 K. This value is by far the highest ever recorded for iron based superconducting wires and has surpassed the threshold for practical application. Also, the hot pressed 122 conductors have superior $J_c$ than MgB$_2$ and NbTi in field region over 10 T. The hot pressing was performed for the 60 mm long *ex-situ* processed Sr-122 tapes in an argon gas flow [37].

The flat rolled mono- and multifilamentary tapes were sandwiched between two pieces of metal sheets and pressed at ~30 MPa at 850-900 ℃ for 30 min.

Figure 10 shows the cross sectional SEM images of Sr-122 cores obtained from the hot pressed and the normally heat treated tapes, respectively [20]. Clearly less voids and the increase of core density were observed in the hot pressed tape. On the other hand, the presence of uniformly textured grains along the tape axis is clearly observed from figure 10(c). Furthermore, as shown in the enlarged part in figure 10(c), grain bending can be observed throughout the pressed sample, which is thought to be caused by the softness of grains during hot deformation. The bending of grains can prevent the formation of cracks and improve the grain coupling effectively, inducing substantial enhancement in grain linkages. As a result, in hot pressed Sr-122 samples, we did not observe residual cracks as in the cold pressed Ba-122 tapes [19, 36], suggesting that hot deformation can help in eliminating cracks effectively while densifying the superconducting cores. The grain boundary structure was further investigated on atomic scale using high resolution TEM, and the results showed that the Sr-122 phase inside the tapes had many clean grain boundaries with low misorientation angle. Therefore, the superior $J_c$ in these hot pressed Sr-122 tapes can be attributed to higher densification with fewer cracks and improved grain texture.

More importantly, our results about the correlations between the $J_c$ and the hot pressing temperature clearly demonstrated that the grain texture has a strong influence on the transport properties. As shown in figure 11a, the $J_c$ of hot pressed Sr-122 tapes increases systematically as the hot pressing temperature increases from 850 to 900 ℃, and then decreases rapidly when further increasing temperature to 925 ℃. The best $J_c$ is $1.2 \times 10^5$ A/cm$^2$ at 10 T and 4.2 K for samples sintered at 900 ℃, which is a new record value ever reported to date [38]. The correspondingly overall conductor (engineering) current density, $J_e$, reaches $2.6 \times 10^4$ A/cm$^2$. Note that the above temperature dependence of $J_c$ is qualitatively similar to the texture development with hot pressing temperature as shown in figure 11b. however, the hardness $H_v$ values for all hot-pressed tapes are almost the same around 154.0 in spite of hot pressing temperature (figure 11d), suggesting that the core density seems only one of main origins of the $J_c$ increase.

Therefore, significant enhancements in $J_c$ in hot pressed Sr-122 tapes is believed to stem mainly from an improvement of grain connectivity by either reducing weak links or increasing the core density.

**5. Fabrication of practical 122 pnictide wires**

For applications, besides high $J_c$, there are many other performance requirements for conductors. Superconducting conductors for large-scale applications are multifilamentary round or tape-shaped wires in which one or more superconducting filaments are embedded in a matrix of normal metal, such as Cu or Ag, which provides protection against magnetic flux jumps and thermal quenching [39]. Such wires must have sufficient strength to withstand the fabrication process, coil winding, the thermal stresses of cooldown, and operational electromagnetic stresses. They must be capable of carrying operating currents, dc or ac, from hundreds to thousands of amperes at costs comparable to Cu. Engineering current density $J_e$ for operation conditions in applications must attain $>10^4$ A/cm$^2$ in magnetic fields [40]. In particular, the minimum critical tensile stress before loss of critical current density should be in the range of 100 MPa or higher, and minimum tensile, compressive, and bending strains before degradation must be several tenths of a percent. The most significant thing is that superconducting conductor cost must be lower enough, in order to compete against copper in applications.

**5.1 Multifilamentary 122 wires**

Towards practical applications of iron-based superconductors, fabricating multifilamentary wires and tapes is an indispensable step. The first production of multifilamentary 122 type wires has been report by Yao et al. [41]. They fabricated Fe/Ag sheathed Sr-122 seven-core superconducting wires and tapes by the ex-situ PIT method. The reacted Sr-122 powder was packed into Ag tubes (OD: 8 mm and ID: 5 mm), which were drawn into wires of 1.58 mm diameter and cut into short wires. Seven short Ag sheathed wires were then bundled into a Fe tube (OD: 8 mm and ID: 5 mm) and drawn into wires. Multifilamentary wires with 2.00 mm diameter were successively rolled into tapes with different thickness, and annealed at 900 °C for 0.5 h.

In contrast to drawing, flat rolling can efficiently increase the mass density of the superconducting core thus significantly improving the transport $J_c$ of the as-drawn wires. As shown in figure 12, the transport $J_c$ of the best 7-core samples achieved 21.1 kA/cm$^2$ at 4.2 K in self field (Blue curve), and showed very weak magnetic field dependence at high fields. After employing a high quality precursor prepared by optimizing the sintering time, our very recent data of 7-filament Sr-122/Ag/Fe composites are plotted in figure 12 (Red curve), showing a substantial increase of $J_c$ value up to 14 kA/cm$^2$ at 10 T, which is mainly due to a higher connectivity. This result is promising for further improvements in 122 type conductors using high strength composite sheaths.

On the other hand, for multifilamentary 122/Ag wires and tapes it is found that hot pressing can also greatly improve their $J_c$ performance. At present, the highest $J_c$ values reached 6.1×10$^4$ A/cm$^2$ and 3.5×10$^4$ A/cm$^2$ at 4.2 K and 10 T, respectively for hot pressed 7- and 19-core Sr-122 tapes [20]. These values are the highest in multifilamentary pnictide wires at present, further demonstrating the great potential of iron-pnictide conductors for high-field applications. Although the in-field $J_c$ property of the pressed multifilaments is higher than that of the rolled multifilaments due to the high tape densification, as can be seen from figure 13, the pressing technique (cold or hot pressing) can only produce tapes of a few cm in length and is not suitable for industrial length, therefore, there are compelling reasons to further pursue a continuous deformation process, e.g. the rolling technique next step, because rolling is a scaleable and practical process for fabricating long and continuous lengths of pnictide wires.

## 5.2 Mechanical property

Pnictide superconductors showed weak decrease of critical current with increasing the external magnetic field [20]. From the viewpoint of application in high magnetic field, where high electromagnetic forces are present, the conductor strength and its tolerance to mechanical load become the important issue. We reported the first results on electro-mechanical properties of Sr-122/Ag conductors made by *ex-situ* PIT process [42]. Sr-122 tapes were stressed by axial tension at 4.2 K under external fields at 5 - 6 T, and critical currents ($I_c$) have been simultaneously measured. As shown in figure 14, while the critical current of Bi-2212/Ag starts to decrease at ε ~ 0.1%, $I_c$ of Sr-122/Ag tape increases up to ε ~ 0.2% then starts to decrease and further rapidly degrades at the strain above 0.25% [42]. The irreversible strain is still low for applications, but quite comparable to Bi-2212/Ag conductors Dr. Kovac measured for comparison. This is not surprised because the annealed Ag sheath is too soft.

In spite of the present low tolerance of the pure Ag-sheathed Sr-122 samples to mechanical stresses, high field coil windings, similar to those using Bi-2212/Ag conductors, are highly expected. Therefore, the next work in the development of 122 tapes will be the improvement of mechanical strength by employing combination of Ag allowing high transport currents with some Ag- and Fe-alloys (e.g. AgMg, stainless steel) acting as mechanical reinforcement, and thermal stability by optimized deformation sequences yielding finer filaments. Recently, in order to reinforce 122/Ag conductors, Gao et al. attempted to fabricate stainless steel (SS)/Ag double sheathed Ba-122 tapes. After rolling and then heat treating the double-sheath architecture, their transport $J_c$ maintained ~$7.7 \times 10^4$ Acm$^{-2}$ (4.2 K, 10 T) [43], suggesting that Ag-based double composite sheaths are good candidate to improving the mechanical properties of pnictide conductors.

On the other hand, copper sheath is the first choice for manufacturing superconducting wire conductors because of its high electrical and thermal conductivities, low cost and good mechanical properties. More recently, Lin et al. from our group fabricated high-$J_c$ Cu sheathed Sr-122 tape conductors utilizing hot pressing technique. This fabrication process can effectively thwart the diffusion of Cu into polycrystalline Sr-122 core. As a consequence, we achieved high transport $J_c$ of $3.1 \times 10^4$ A/cm$^2$ and high engineering $J_e$ of $10^4$ A/cm$^2$ in 10 T at 4.2 K [44], as shown in figure 15. It should be noted that the $J_e$ has reached the widely accepted value of $10^4$ A/cm$^2$ at 10 T needed for practical applications [40], demonstrating the potential of Cu sheath for practical application of pnictide wires and tapes.

**5.3 Scale up of the first 11 m long-length tape**

High $J_c$ of $10^5$ A/cm$^2$ at 4.2 K has been produced by several groups, such as Florida, NIMS, and the IEE group which is where the author is from. We can now routinely obtain critical current density of $10^5$ A/cm$^2$ at 10 T for short tapes. However, all applications require not only high $J_c$ conductors, but sufficient long length for winding the coils, or cabling as well. In particular, an easy and simple process is required to balance the high performance and the production cost of the long length wires.

In August, 2014, the author's group announced the world first 11m long Sr-122/Ag tape by rolling [45], which is a scalable process that can be used to manufacture km of 122 flat tape that will be needed for practical applications, as shown in figure 16. Transport $J_c$ at 4.2 K and 10 T measured along the length of the Sr-122 tape is shown in figure 17. It exhibits fairly uniform $J_c$ distribution throughout the tape, the maximum and minimum $J_c$ were ~$2.12 \times 10^4$ and ~$1.68 \times 10^4$ A/cm$^2$ at 10 T, 4.2 K, respectively. Clearly, an average $J_c$ value of ~$1.84 \times 10^4$ A/cm$^2$ at 10 T was achieved over the 11 meter length, demonstrating the high performance of long 122 tapes made by the scalable rolling process.

For high magnetic field applications, fabricating longer multifilament 122 wires is indispensable. There is a high likelihood that 10-100 meter lengths of 122 conductor will be available for applications several years later.

## 6. Other type wires

Wires and tapes of the 1111 family are appealing for applications as well due to its highest $T_c$ among iron-based superconductors. However, less work has been carried out for the fabrication of 1111 wires and tapes, because of the difficulty in controlling O and F stoichiometry during heat treatments at high temperatures. Recently, Zhang et al. synthesized Fe-sheathed SmFeAsO$_{0.8}$F$_{0.2}$ tapes using a Sn-presintering process and achieved a high $J_c$ of 34.5 kA/cm$^2$ at 4.2 K in the self field [46]. However, $J_c$ rapidly decreased with increasing the magnetic field when compared to 122 tapes, possibly because of relatively more impurities in the superconducting core caused by the severe fluorine loss during the final sintering step in this system.

More recently, the Sefat group reported the first wire fabrication of non-arsenic Ba(NH$_3$)Fe$_2$Se$_2$ iron-based superconductor by the PIT method [47]. Ba(NH$_3$)Fe$_2$Se$_2$ has lower toxicity compared with arsenic 122 iron-based superconductors (e.g. Ba$_{0.6}$K$_{0.4}$Fe$_2$As$_2$). Their non-arsenic 122 wires with $T_c$=16 K showed the transport critical current densities of $10^3$ A/cm² at 4.2 K, although bulk materials are estimated to carry $J_c$ >$10^5$ A/cm² (4 K, self-field).

For the 11 type iron chalcogenide compounds, represented by FeSe and Fe(Se, Te), the latest progress was reported by Palomno et al. [48]. They made an extensive research on the sheath materials for ex-situ PIT Fe(Se, Te) wires, and confirmed Fe sheath to be most chemically compatible with Fe(Se,Te) phase. However, the compositional modification and phase de-mix induced by Fe sheath during heat treatment still limited the transport Jc on the level of several hundred A/cm² at 4.2 K in self-field, which is comparable to the previously reported ex-situ PIT 11 type wires.

## 7. Summary and outlook

In summary, ex-situ PIT method is now the most popular method to fabricate iron-based superconducting tapes and wires. As sheath materials, so far Ag shows no reaction with 122 phase, but Fe and Cu appear as a possible alternative, in particular if the final heat treatment temperature can be further reduced. Main problems encountered when producing 122 wires or tapes by the PIT method are the need for an improvement of grain connectivity, by either a densification of the filaments or a texturing process. In the past several years, rapid progress on 122 wire with a view towards its development as a technical conductor has reached remarkable milestones. The critical current results of figure 18 represent the improvement of 122 type conductors made using a PIT process mainly by three groups of IEECAS, NIMS and Florida. The best $J_c$ performance has now reached the 120 kA/cm² level (10 T and 4.2 K) for hot pressed 122/Ag composites, corresponding to overall or engineering critical current densities $J_e$ of 26 kA/cm². If using the Cu instead of Ag as sheath material, the 122/Cu tapes have recorded a high $J_c$ of above 31 kA/cm² at 4.2 K under 10 T (Correspondingly, $J_e$=10 kA/cm²). On the other hand, for multifilamentary 122/Ag wires and tapes, the highest $J_c$ values of 61 kA/cm² and 35 kA/cm² at 4.2 K and 10 T are obtained for 7- and 19-core tapes, respectively. Using the scalable rolling process, the 11 m long 122/Ag tapes with an average $J_c$ exceeding 18.4 kA/cm² at 10 T were produced for the first time by the IEE group. These results demonstrate that the electrical performance of commercially interesting multifilamentary and long length wires has established parity with that of monofilamentary samples, making this already fascinating 122 superconductor of even greater interest for applications.

It seems clear that for 122 wires the improvement of $J_c$ can still be obtained by further optimization of the temperatures, annealing times, and deformation conditions as well as enhancement in flux pinning in high fields.

Now, the main challenge is how to transfer the already achieved high $J_c$ to the large scale required for application. At the same time, thermal stability and mechanical properties must be improved. With what has been achieved, it is believed that PIT processed 122 type wires have the strong potential to be produced industrially in long lengths for high-field magnets.

**Acknowledgments**

We would like to thank our colleagues and students: Xianping Zhang, Chao Yao, He Lin, Chiheng Dong, Dongliang Wang, Qianjun Zhang, He Huang for their work on pnictides. We wish to thank our collaborators, whose published works are quoted in this paper. This work is partially supported by the National '973' Program (grant No. 2011CBA00105), the National Natural Science Foundation of China (51320105015), the Beijing Municipal Science and Technology Commission (No. Z141100004214002), and the Beijing Traning Project For the Leading Talents in S & T under Grant No. Z151100000315001.

**References**

[1] Y. Kamihara, T. Watanabe, M. Hirano, H. Hosono, J. Am. Chem. Soc. 130 (2008) 3296.

[2] X.L. Wang, S.R. Ghorbani, G. Peleckis, S.X. Dou, Adv. Mater. 21 (2009) 236.

[3] K. Iida, J. Hanisch, C. Tarantini, F. Kurth, J. Jaroszynski, S. Ueda, M. Naito, A. Ichinose, I. Tsukada, E. Reich, V. Grinenko, L. Schultz, B. Holzapfel, Sci. Rep. 3 (2013) 2139.

[4] K. Tanabe, H. Hosono, Jpn. J. Appl. Phys. 51 (2012) 010005.

[5] T. Katase, Y. Ishimaru, A. Tsukamoto, H. Hiramatsu, T. Kamiya, K. Tanabe, H. Hosono, Nat. Commun. 2 (2011) 409.

[6] Y. W. Ma, Supercond. Sci. Technol. 25 (2012) 113001.

[7] Z.S. Gao, L. Wang, Y.P. Qi, D.L. Wang, X.P. Zhang, Y.W. Ma, Supercond. Sci. Technol. 21 (2008) 105024.

[8] Y. P. Qi, X. P. Zhang, Z. S. Gao, Z. Y. Zhang, L. Wang, D. L. Wang, Y. W. Ma, Physica C 469 (2009) 717.

[9] X.P. Zhang, L. Wang, Y.P. Qi, D.L. Wang, Z.S. Gao, Z.Y. Zhang, Y.W. Ma, Physica C 470 (2010) 104.

[10] K. Togano, A. Matsumoto, H. Kumakura, Supercond. Sci. Technol. 23 (2010) 045009.

[11] Y.P. Qi, L. Wang, D.L. Wang, Z.Y. Zhang, Z.S. Gao, X.P. Zhang, Y.W. Ma, Supercond. Sci. Technol. 23 (2010) 055009.

[12] L. Wang, Y.P. Qi, Z.S. Gao, D.L. Wang, X.P. Zhang, Y.W. Ma, Supercond. Sci. Technol. 23 (2010) 025027.

[13] C. Yao, C.L. Wang, X.P. Zhang, L. Wang, Z.S. Gao, D.L. Wang, C.D. Wang, Y.P. Qi, Y.W. Ma, S. Awaji, K. Watanabe, Supercond. Sci. Technol. 25 (2012) 035020.

[14] L. Wang, Y.P. Qi, X.P. Zhang, D.L. Wang, Z.S. Gao, C.L. Wang, C. Yao, Y.W. Ma, Physica C 471 (2011) 1689.

[15] Z.S. Gao, Y.W. Ma, C. Yao, X.P. Zhang, C.L. Wang, D.L. Wang, S. Awaji, K. Watanabe, Sci. Rep. 2 (2012) 998.

[16] J.D. Weiss, C. Tarantini, J. Jiang, F. Kametani, A.A. Polyanskii, D.C. Larbalestier, E.E. Hellstrom, Nat. Mater. 11 (2012) 682.

[17] K. Togano, Z.S. Gao, H. Taira, S. Ishida, K. Kihou, A. Iyo, H. Eisaki, A. Matsumoto, H. Kumakura, Supercond. Sci. Technol. 26 (2013) 065003.

[18] C. Yao, H. Lin, X.P. Zhang, D.L. Wang, Q.J. Zhang, Y.W. Ma, S. Awaji, K. Watanabe, Supercond. Sci. Technol. 26 (2013) 075003.

[19] Z. Gao, K. Togano, A. Matsumoto, H. Kumakura, Sci. Rep. 4 (2014) 4065.

[20] X.P. Zhang, C. Yao, H. Lin, Y. Cai, Z. Chen, J.Q. Li, C.H. Dong, Q.J. Zhang, D.L. Wang, Y.W. Ma, H. Oguro, S. Awaji, K. Watanabe, Appl. Phys. Lett. 104 (2014) 202601.

[21] C.L. Wang, L. Wang, Z.S. Gao, C. Yao, D.L. Wang, Y.P. Qi, X.P. Zhang, Y.W. Ma, Appl. Phys. Lett. 98 (2011) 042508.

[22] H. Lin, C. Yao, X.P. Zhang, H.T. Zhang, D.L. Wang, Q.J. Zhang, Y.W. Ma, Physica C 495 (2013) 48.

[23] L. Wang, Y.P. Qi, D.L. Wang, X.P. Zhang, Z.S. Gao, Z.Y. Zhang, Y.W. Ma, S. Awaji, G. Nishijima, K. Watanabe, Physica C 470 (2010) 183.

[24] K. Togano, A. Matsumoto, H. Kumakura, Appl. Phys. Express. 4 (2011) 043101.

[25] Y.W. Ma, L. Wang, Y.P. Qi, Z.S. Gao, D.L. Wang, X.P. Zhang, IEEE Trans. Appl. Supercond. 21 (2011) 2878.

[26] M. Putti et al., Supercond. Sci. Technol. 23 (2010) 034003.

[27] J.H. Durrell, C.B. Eom, A. Gurevich, E.E. Hellstrom, C. Tarantini, A. Yamamoto, D.C. Larbalestier, Rep. Prog. Phys. 74 (2011) 124511.

[28] R. Flukiger, T. Graf, M. Decroux, C. Groth, Y. Yamada, IEEE Trans. Magn. 27 (1991) 1258.

[29] Y.W. Ma, C. Yao, X.P. Zhang, H. Lin, D.L. Wang, A. Matsumoto, H. Kumakura, Y. Tsuchiya, Y. Sun, T. Tamegai, Supercond. Sci. Technol. 26 (2013) 035011.

[30] C.H. Dong, C. Yao, H. Lin, X.P. Zhang, Q.J. Zhang, D.L. Wang, Y.W. Ma, H. Oguro, S. Awaji, K. Watanabe, Scr. Mater. 99 (2015) 33.

[31] F.K. Lotgering, J. Inorg. Nucl. Chem. 9 (1959) 113.

[32] R. Flukiger, H.L. Suo, N. Musolino, C. Beneduce, P. Toulemonde, P. Lezza, Physica C 385 (2003) 286.

[33] S. Pyon, Y. Tsuchiya, H. Inoue, H. Kajitani, N. Koizumi, S. Awaji, K. Watanabe, T. Tamegai, Supercond. Sci. Technol. 27 (2014) 095002.

[34] Q. Li, K. Brodersen, H.A. Hjuler, T. Freltoft, Physica C 217 (1993) 360.

[35] G. Grasso, A. Jeremie, R. Flukiger, Supercond. Sci. Technol. 8 (1995) 827.

[36] K. Togano, Z.S. Gao, A. Matsumoto, H. Kumakura, Supercond. Sci. Technol. 26 (2013) 115007.

[37] H. Lin, C. Yao, X.P. Zhang, H.T. Zhang, D.L. Wang, Q.J. Zhang, Y.W. Ma, S. Awaji, K. Watanabe, Sci. Rep. 4 (2014) 4465.

[38] H. Lin, C. Yao, X.P. Zhang, C.H. Dong, H.T. Zhang, D.L. Wang, Q.J. Zhang, Y.W. Ma, S. Awaji, K. Watanabe, H.F. Tian, J.Q. Li, Sci. Rep. 4 (2014) 6944.

[39] D. Larbalestier, A. Gurevich, D.M. Feldmann, A. Polyanskii, Nature 414 (2001) 368.

[40] J. Shimoyama, Supercond. Sci. Technol. 27 (2014) 044002.

[41] C. Yao, Y.W. Ma, X.P. Zhang, D.L. Wang, C.L. Wang, H. Lin, Q.J. Zhang, Appl. Phys. Lett. 102 (2013) 082602.

[42] P. Kovac, L. Kopera, T. Melisek, M. Kulich, I. Husek, H. Lin, C. Yao, X. Zhang, Y. Ma, Supercond. Sci. Technol. 28 (2015) 035007.

[43] Z. Gao, K. Togano, A. Matsumoto, H. Kumakura, Supercond. Sci. Technol. 28 (2015) 012001.

[44] He Lin, Chao Yao, Xianping Zhang, Chiheng Dong, Haitao Zhang, Dongliang Wang, Qianjun Zhang, Yanwei Ma, 2015 arXiv:1504.02219.

[45] Xianping Zhang, He Lin, Chao Yao, Yanwei Ma, Invited presentation at the 2014 Applied Superconductivity Conference, August 10-15, 2014, Charlotte, USA.

[46] Q.J. Zhang, C. Yao, H. Lin, X.P. Zhang, D.L. Wang, C.H. Dong, P.S. Yuan, S.P. Tang, Y.W. Ma, S. Awaji, K. Watanabe, Y. Tsuchiya, T. Tamegai, Appl. Phys. Lett. 104 (2014) 172601.

[47] J. E. Mitchell, D. A. Hillesheim, C. A. Bridges, M. P. Paranthaman, K. Gofryk, M. Rindfleisch, M. Tomsic, A. S. Sefat, Supercond. Sci. Technol. 28 (2015) 045018.

[48] M. Palombo, A. Malagoli, M. Pani, C. Bernini, P. Manfrinetti, A. Palenzona, M. Putti, J. Appl. Phys. 117 (2015) 213903

# Captions

Figure 1 Fabrication of iron-based wires by powder-in-tube process.

Figure 2 $J_c$ at 4.2K as a function of magnetic field for Sn added Sr-122 tapes heat treated at different temperatures. [22]

Figure 3 $J_c$ as a function of the applied magnetic field at 4.2 K for Sr-122/Ag samples with various tape thickness. Note that the data of ϕ1.9 mm round wire is shown for comparison.

Figure 4 $J_c$ (4.2 K, 10 T) as a function of the tape thickness for Fe-sheathed samples.

Figure 5 SEM image of cross section of a rolled Sr-122 filament after annealing. The arrow indicates the rolling direction. [14]

Figure 6 Magnetic field dependence of the transport $J_c$ at various measuring temperatures for the Sr-122/Fe tapes. [29]

Figure 7 $J_c$ at 4.2 K as a function of applied field for the rolled Fe-sheathed Sr-122 and Ag-sheathed Ba-122 tapes.

Figure 8 Transport $J_c$ at 4.2 K as a function of applied field for Ag-sheathed 122 tapes fabricated at different thermomechanical processes. Roll means rolling, HT means heat treatment, Press means cold pressing.

Figure 9 Comparison between the variation of $J_c$ *vs* applied field at 4.2 K for the rolled and hot pressed Sr-122/Ag tapes. [20]

Figure 10 SEM images of the core region of Sr-122 tapes. (a) Flat rolled samples viewed from the tape surface direction. Some voids and cracks were indicated by arrows or circles. (b) Pressed samples viewed from the tape surface direction. (c) Pressed samples viewed from the longitudinal cross-sections. [20]

Figure 11 The hot pressing (HP) temperature dependence of transport $J_c$ for Sr-122/Ag tapes. (a), c-axis texture parameter *F* (b), residual resistivity ratio *RRR* (c) and average Vickers micro-hardness *Hv* values (d). [38]

Figure 12 The field dependence of transport $J_c$ at 4.2 K for 7-multifilamentary Fe-Ag composite sheathed Sr-122 tapes (thickness: 1.00, 0.80, and 0.60mm).

Figure 13 Variation of $J_c$ *vs* applied field at 4.2 K for the rolled and hot pressed 19-multifilamentary Sr-122/Ag tapes. Inset shows SEM image of the cross section of rolled 19-core tape.

Figure 14 Normalized critical current versus axial strain for Ag-cladded Sr-122 tapes.

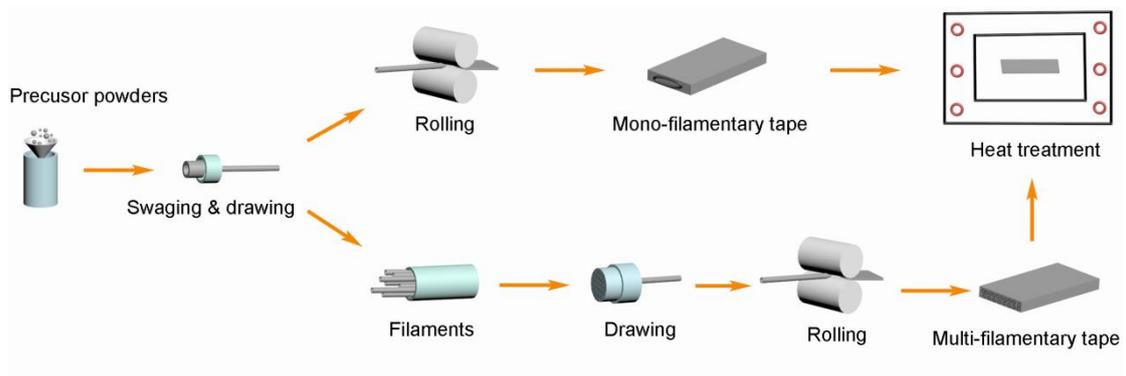

Figure 1

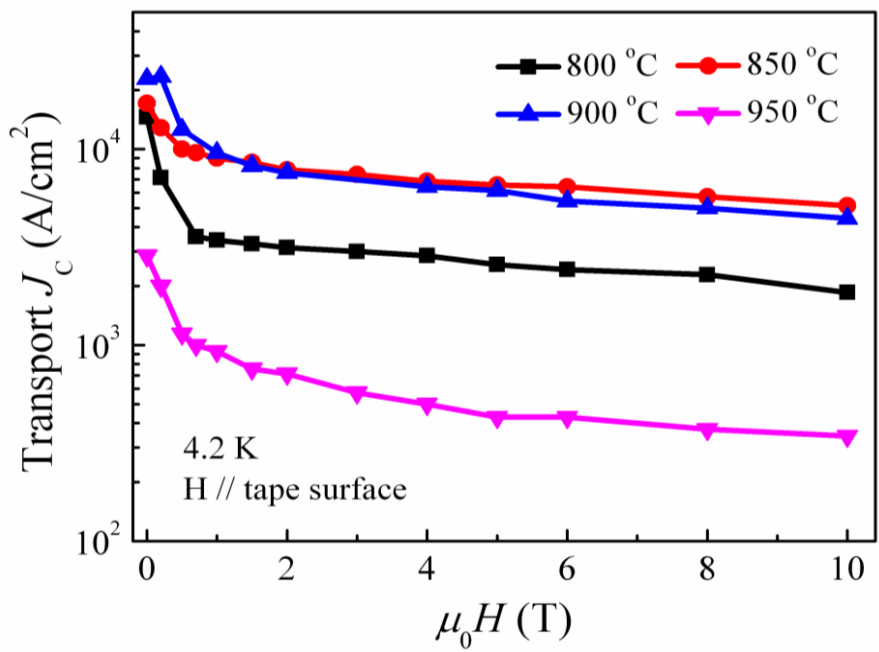

Figure 2

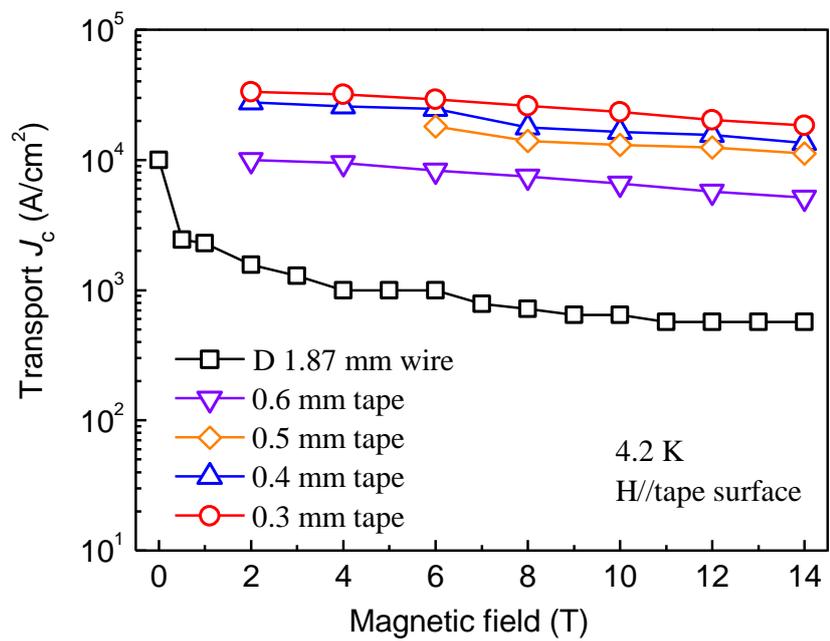

Figure 3

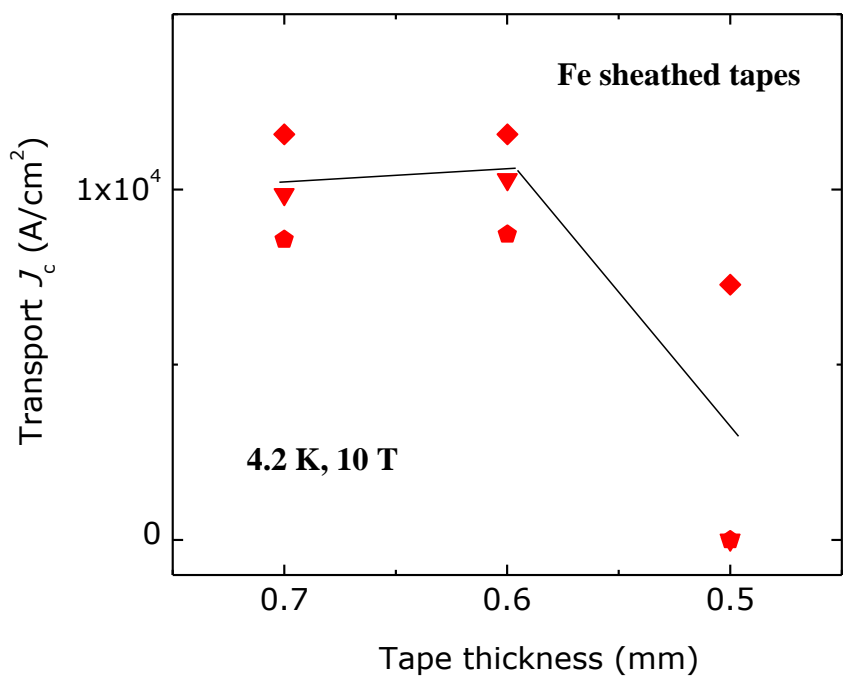

Figure 4

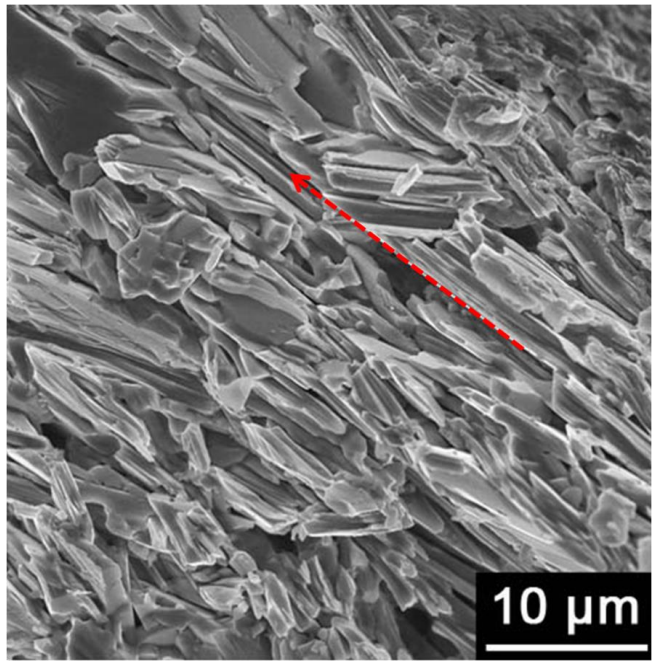

Figure 5

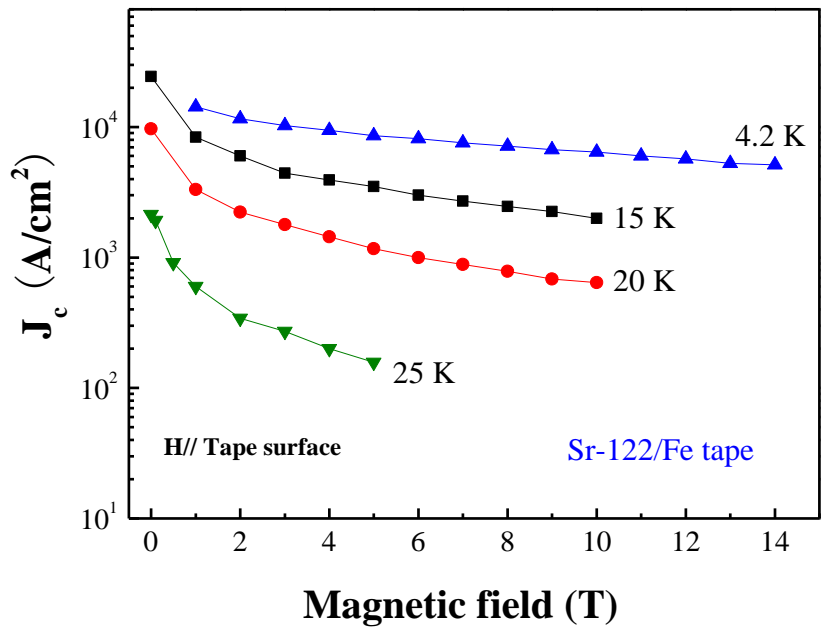

Figure 6

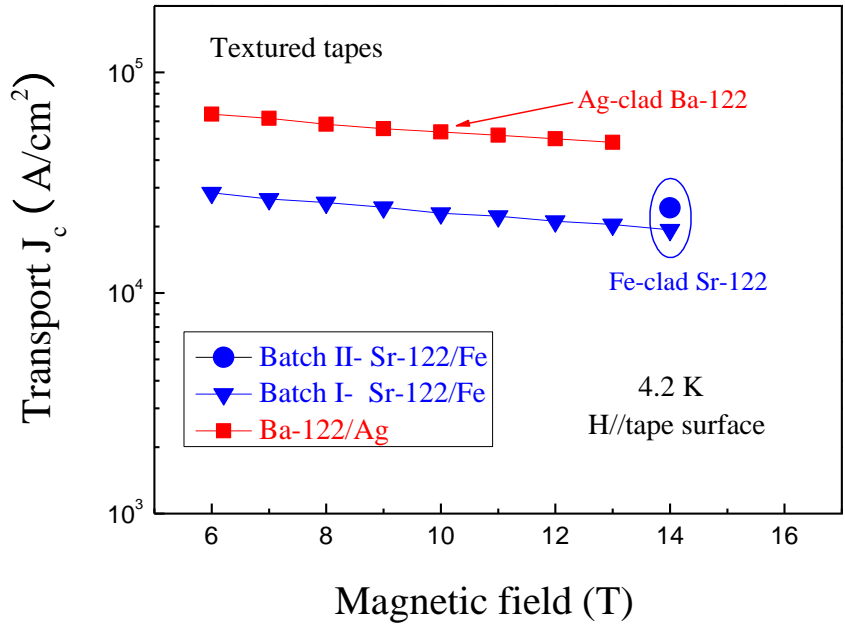

Figure 7

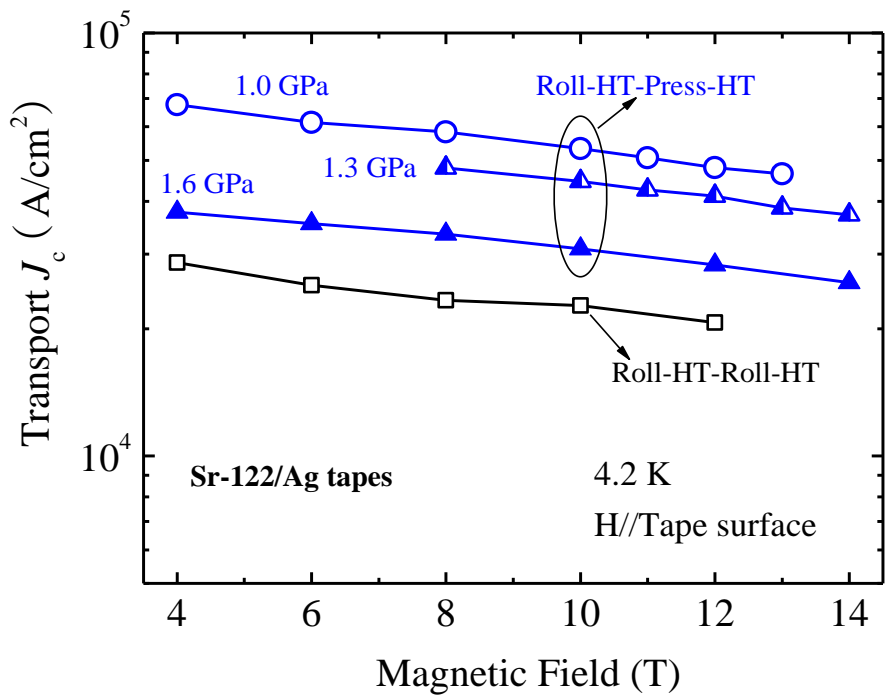

Figure 8

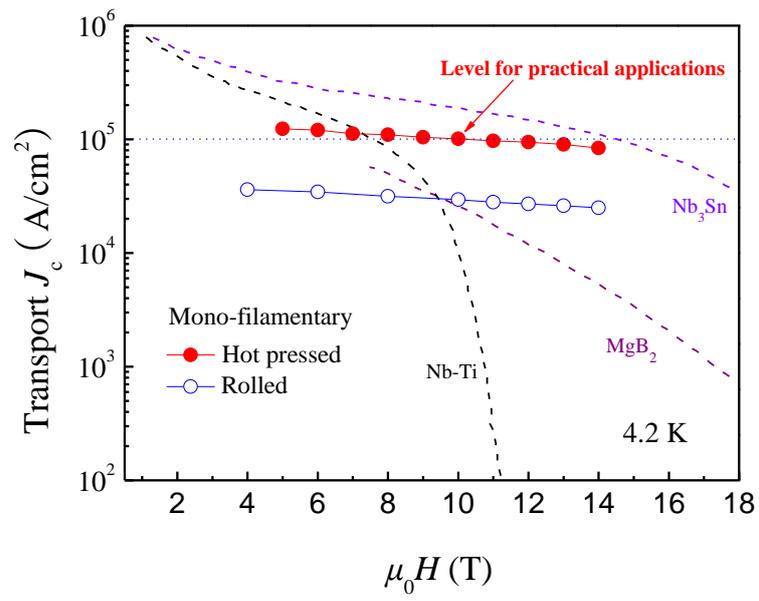

Figure 9

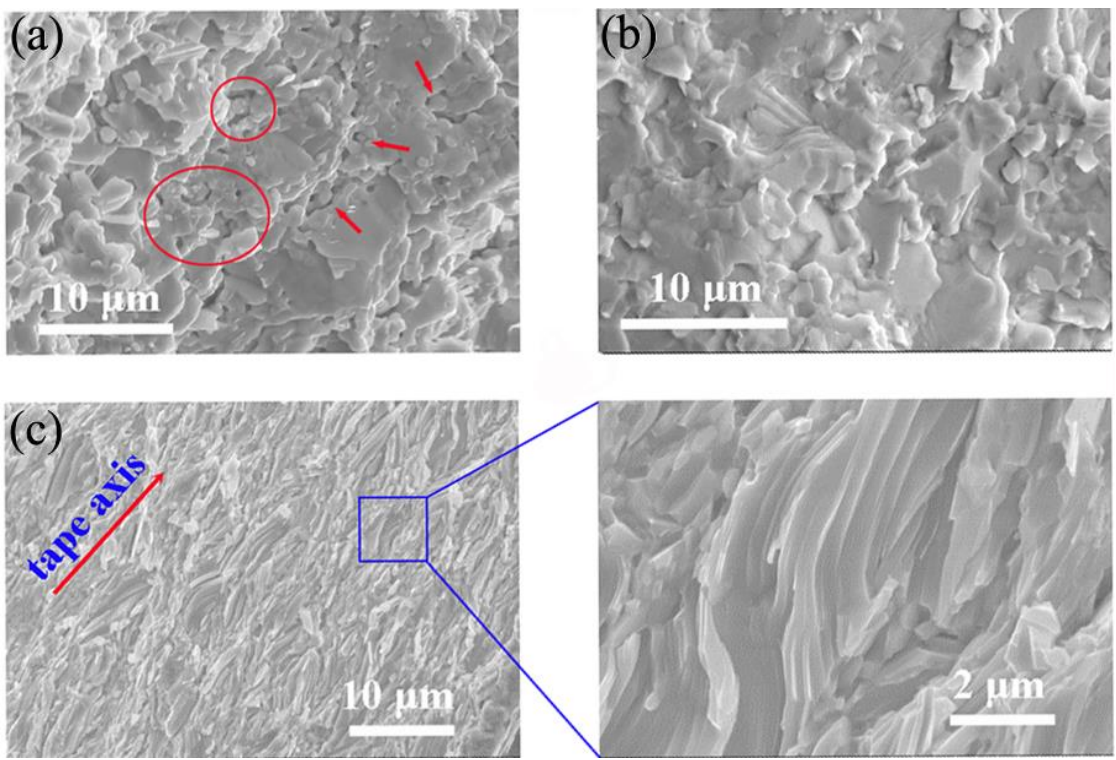

Figure 10

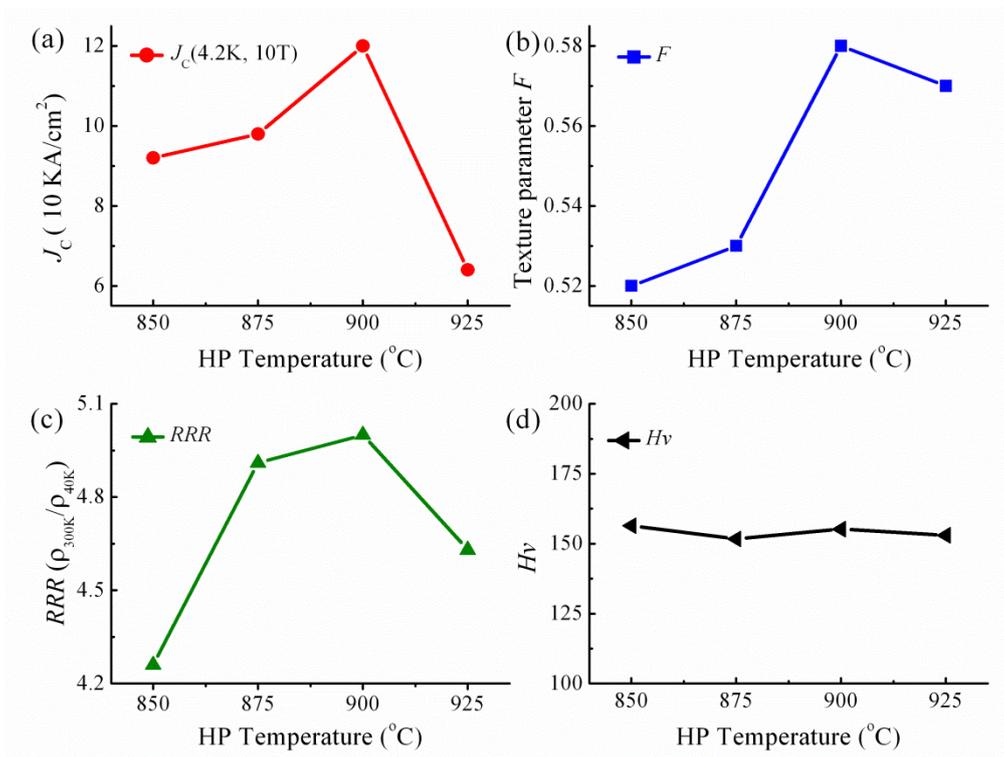

Figure 11

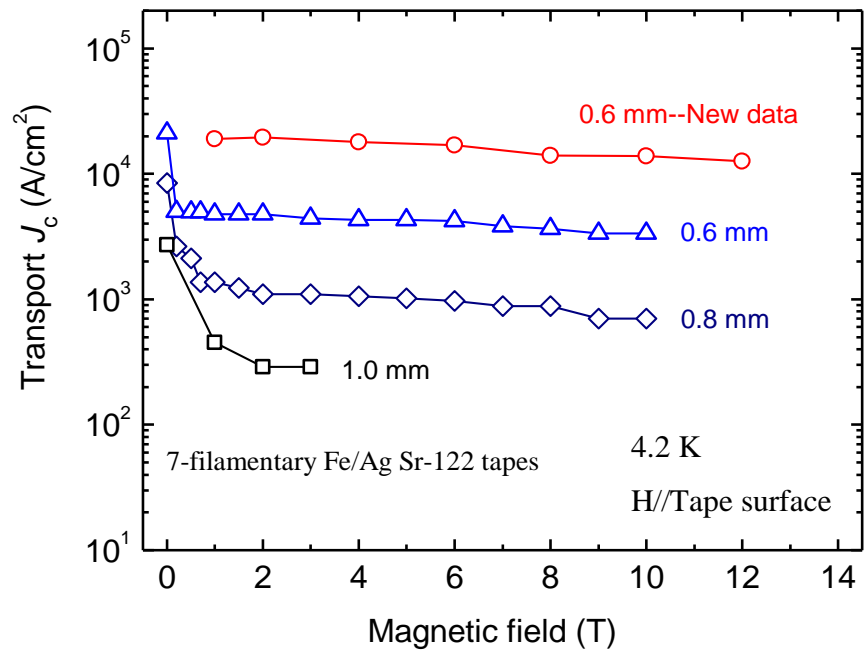

Figure 12

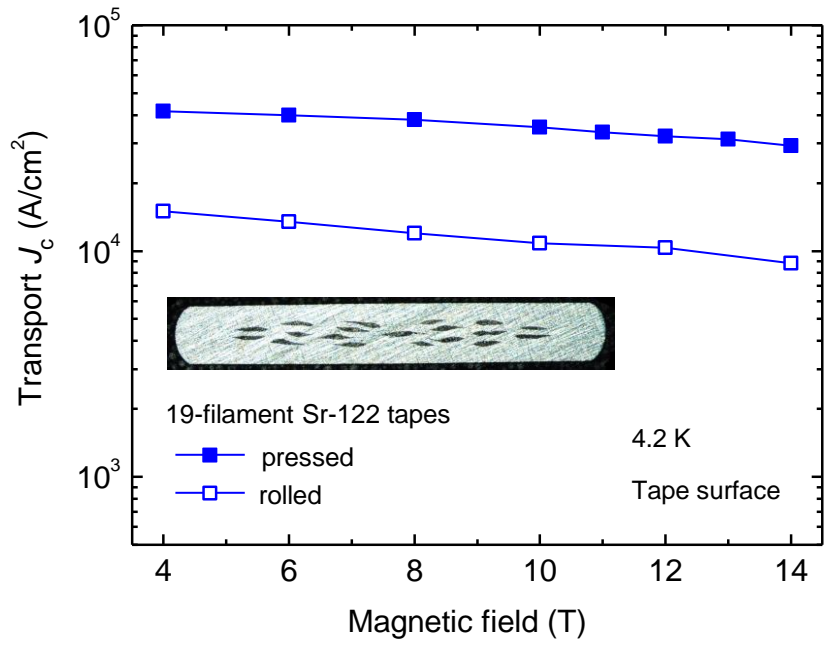

Figure 13

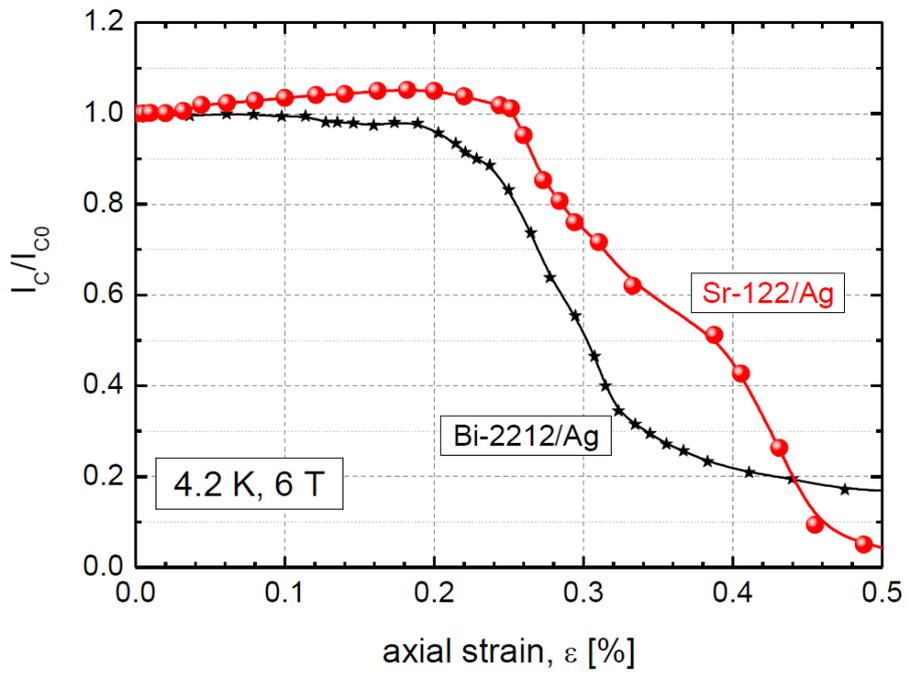

Figure 14

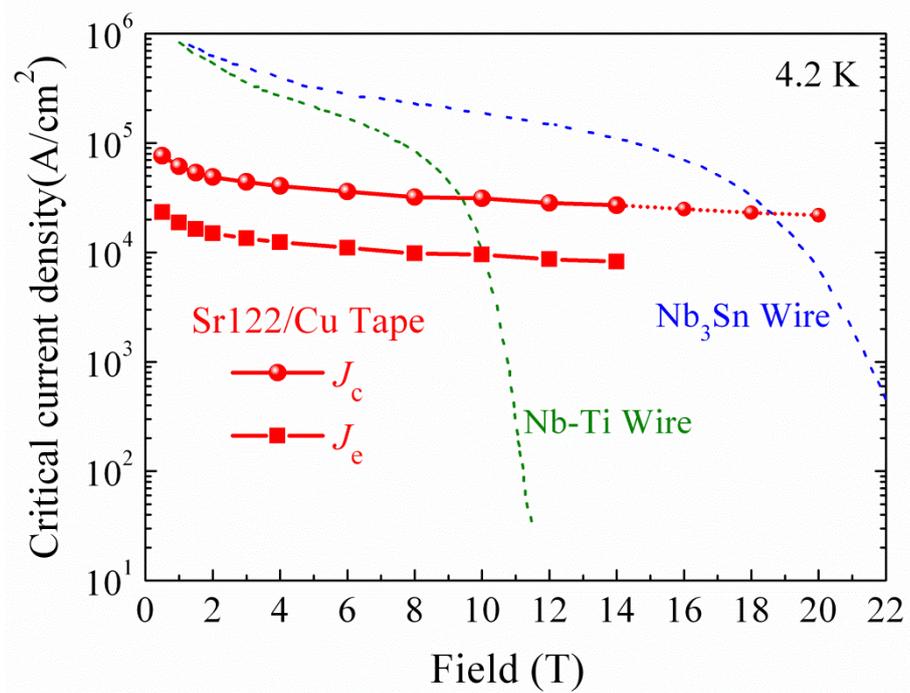

Figure 15

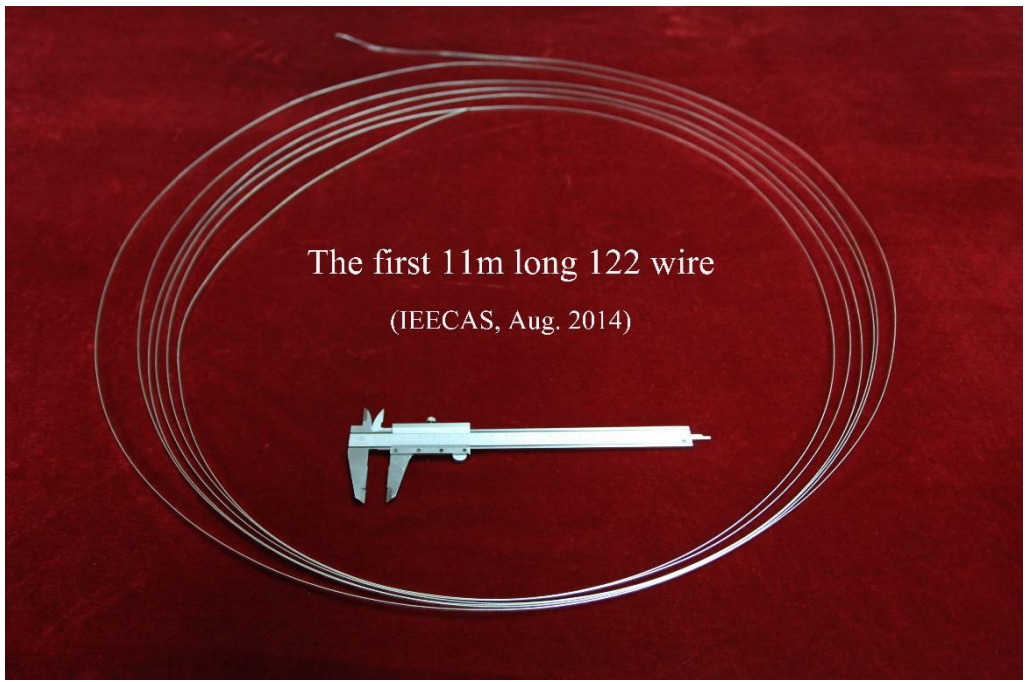

Figure 16

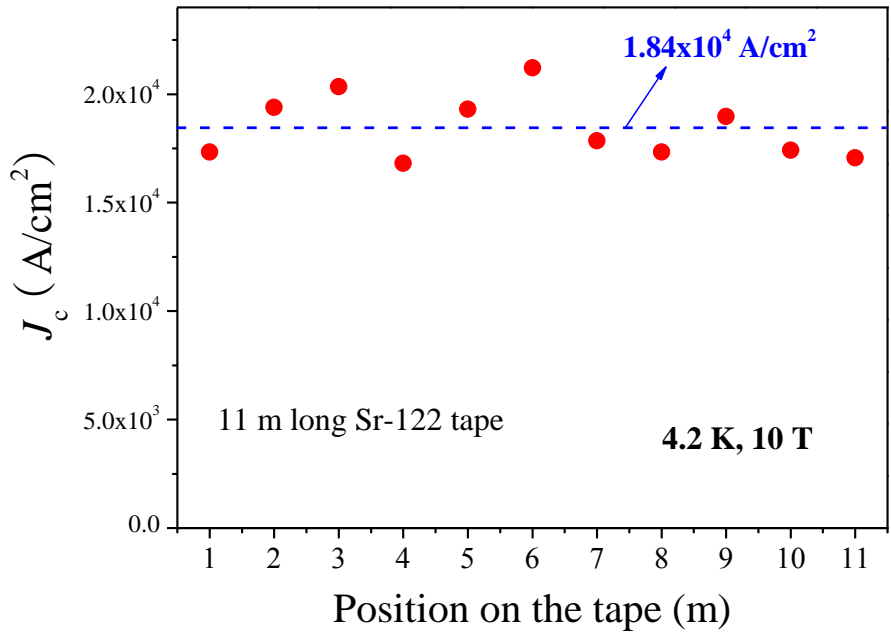

Figure 17

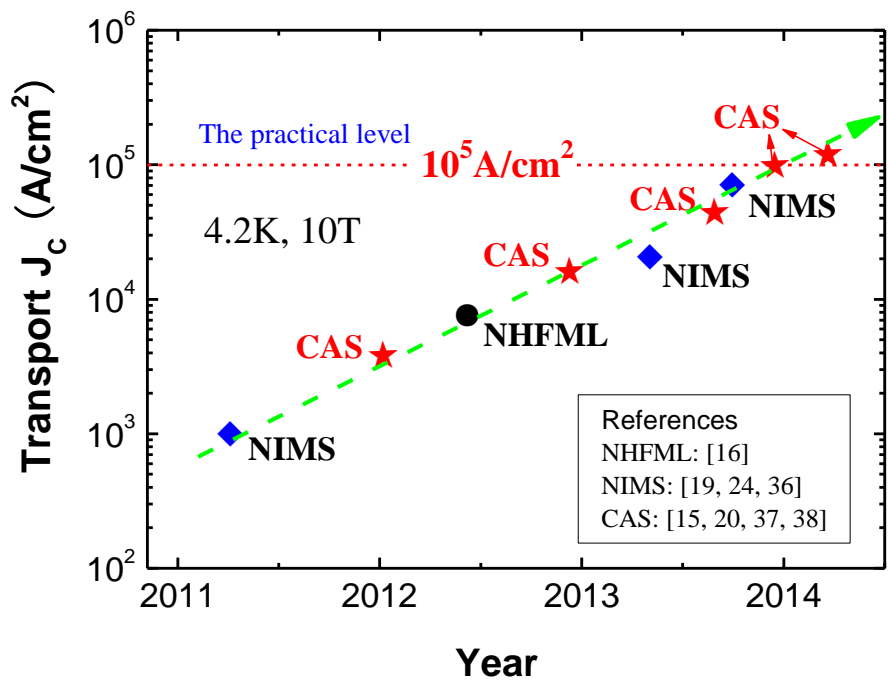

Figure 18